\documentclass[aip,jcp,reprint]{revtex4-1}
\usepackage{graphicx}
\usepackage[colorlinks]{hyperref}
\usepackage{amsmath,amssymb}
\usepackage{chemformula}
\begin{document}
	\title{Charge Regulation of Colloidal Particles: Theory and Simulations}
	
	\author{Amin Bakhshandeh}
	\email{amin.bakhshandeh@ufrgs.br}
	\affiliation{Programa de P\'os-Gradua\c c\~  ao em F\'\i sica, Instituto de F\'\i sica e Matem\'{a}tica, Universidade Federal de Pelotas, Caixa Postal 354, CEP 96010-900 Pelotas, RS, Brazil}
	\author{Derek Frydel} 
	\email{derek.frydel@usm.cl}
	\affiliation{Department of Chemistry, Federico Santa Maria Technical University, Campus San Joaquin, Santiago, Chile}
	\author{Alexandre Diehl} 
	\email{diehl@ufpel.edu.br}
	\affiliation{Departamento de F\'\i sica, Instituto de F\'\i sica e Matem\'{a}tica, Universidade Federal de Pelotas, Caixa Postal 354, CEP 96010-900 Pelotas, RS, Brazil}
	\author{Yan Levin}
	\email{levin@if.ufrgs.br }
	\affiliation{Instituto de F\'isica, Universidade Federal do Rio Grande do Sul, Caixa Postal 15051, CEP 91501-970, Porto Alegre, RS, Brazil}

	\begin{abstract}
		
		  To explore charge regulation (CR) in physicochemical  and biophysical systems, we present a model of colloidal particles with sticky adsorption sites which account for the formation of covalent bonds between the hydronium ions and the surface functional groups. Using this model and  Monte Carlo simulations, we find that the standard Ninham and Parsegian (NP) theory of CR leads to results which deviate significantly from computer simulations.  
The problem of NP approach is traced back to the use of bulk equilibrium constant to account for surface chemical reactions.   To resolve this difficulty we present a new theory of CR.  The fundamental ingredient of the new approach is the sticky length, which is non-trivially related with the bulk equilibrium constant. The theory is found to be in excellent agreement with computer simulations, without any adjustable parameters. As an application of the theory we calculate the effective charge of colloidal particles containing carboxyl groups, as a function of pH and salt concentration.
		
		
	\end{abstract}
	
	\maketitle
	Electrostatic interactions play a fundamental role in physics, chemistry, and biology.  
	The long-range nature of the Coulomb force, however, makes it very difficult to study theoretically~\cite{levin}. 
	In aqueous systems ions are usually hydrated by water molecules.   On the other hand, acids lose proton, which associates with the water molecule forming a hydronium ion~\cite{agmon}.
There are many reactions that are controlled by pH, and the acid-base equilibrium directly influences the functionality of biomolecules.
	Although pH can be easily tuned in experiments, it is much more difficult to account for the chemical equilibrium in theoretical 
	and simulation studies~\cite{baer}. 
	
	Colloidal particles often have organic functional groups on their surfaces.  In aqueous systems 
	these groups dissociate, loosing a proton, resulting in a colloidal surface charge~\cite{pincus,trizac1,trizac2,palberg}.  The amount of  surface charge  strongly depends on the pH of the environment~\cite{adamson,markovich} and is controlled by the chemical equilibrium between hydronium ions and the functional groups.   This process is known as charge regulation (CR)~\cite{podgornik2018,frydel2019,Avni,AVNI2019,sen2019internal,polymer2018,ozcelik2019}.
	The concept of charge regulation was first described  by Linderstr{\o}m-Lang~\cite{Lang,lund2013,gitlin2003} and
	studied theoretically  by Ninham and Parsegian.~\cite{ninham}.
	CR is of fundamental importance in colloidal science~\cite{podgornik2018,prieve1976,carnie1993,
		behrens1999,netz2002,henle2004,trefalt2015,Majee,lovsdorfer2018,hallett,waggett2018,trefalt2015charge,Roij} and biophysics~\cite{lund,lowen1,monica1,monica2, monica3,govrin,henrik}. It has been applied to explore the 
stability of electrical double layers~\cite{podgornik1991,markovich,Markovich_2017,leckband,da2009,smith2018,Podgorni2015} and is of great technological importance in fields as diverse as mineral preparation, agriculture, ceramics, and surface coating~\cite{hunter2001s}.  
	
	Consider a weak acid \ch{HA} in equilibrium with bulk water,
$\ch{HA} +\ch{H_2 O}   \rightleftarrows \ch{H_3 O+}  + \ch{A-}$.
	For dilute solutions the concentration of all species is controlled by the law of mass action, $K_{eq} =c_{\ch{HA} }/c_{\ch{A-}}c_{\ch{H+}}$,
	where $K_{eq}$ is the equilibrium constant and $c$ indicates the concentration of each specie.  Ninham and Parsegian (NP) supposed that the same equilibrium relation will hold for the reactive (acidic) sites on the colloidal surface with the local concentration of hydronium  determined by the Boltzmann distribution, $c_{\ch{H+}}^{surf}=c_{\ch{H+}}^{bulk} \exp({-\beta q \phi_0})$
	where  $\beta=1/k_BT$, $q$ is the proton charge, and $\phi_0$ is the surface electrostatic potential. 
	NP concluded that the effective surface charge of the colloidal particle will be renormalized from its bare value $-q\sigma_0$, corresponding to all functional groups being ionized, by the associated protons.  Taking into account the surface equilibrium of hydronium through the Langmuir adsorption isotherm, they argued that one can use the usual  Poisson-Boltzmann (PB) equation to account for the distribution of ions around the colloidal particle, but with the effective renormalized surface charge given by
	\begin{equation}    
	q \sigma_{r} = -q\sigma_0 +\frac{ K_{eq}N_{site}q~c_a~\mathrm{e}^{-\beta q \phi_0}}{4~\pi~a^2 (1 +K_{eq}~c_a~\mathrm{e}^{-\beta q \phi_0 }) },
	\label{Eq3}
	\end{equation}
	where $N_{site}$ is the number of ionizable surface groups,  $a$ is the colloidal radius, and $c_a$ is the bulk concentration of hydronium ions, $c_a=10^{-pH}$~M . 
	Within the NP formalism $K_{eq}$ is the usual bulk equilibrium constant. If the surface groups are strongly acidic $K_{eq} \rightarrow 0$, all the surface groups are ionized, $q\sigma_{r} \rightarrow -q\sigma_0$.  
	
	NP theory has been extensively used to study various biological and chemical systems.  However, since within the
	experiment there is always uncertainty about the underlying physical parameters --- the surface charge, location of the shear plane, the appropriate equilibrium constant ---it is hard to judge the validity of a theory.  In this Letter we propose a model of CR which has no ambiguities of an experimental system and can be solved exactly using computer simulations.  With the help of this model, we find that predictions of NP theory deviate significantly from the results of simulations. We then introduce a new theory which agrees perfectly with the simulation data, allowing us to uniquely predict the number of ionized groups and the ionic distribution around a CR colloidal particle, without any adjustable parameters.  
	
	The fundamental parameter in the NP theory is the equilibrium constant.  We start, then, by showing how the equilibrium constant can be calculated from a microscopic model.  To do this we first consider the Baxter model~\cite{baxter} of sticky hard spheres of species H and A. For simplicity we will suppose that both particles have the same diameter, $d=2r_{ion}$. The H-H and A-A interactions are purely hardcore repulsion, while collisions between H and A can result in formation of molecules HA.  The H-A interaction potential is $u = u_{st} + u_{ hc}$, where  the hard core potential is
$u_{ hc} = \infty$ for $r<d$ and $0$ otherwise. The attractive sticky potential of range $\delta_r$,
	\begin{equation}  
	\begin{split}
		u_{st}(r) =
	\begin{cases}
	0, & r<d~\text{and}~ r > d+\delta_r,\\
	-\epsilon, &          d \le r \le d+\delta_r,\\
	\end{cases} \\
	\end{split}
	\label{Eq4}  
	\end{equation}  
	is used to model the chemical bonding between H and A. 
	The Boltzmann factor for the sticky potential can then be written as $e^{-\beta u_{st}(r)} = 1 + \delta_r(\mathrm{e}^{\beta \epsilon} -1)\Delta(r) $,
where $\Delta(r)$ is~\cite{frydel2019}  
	\begin{equation}  
	\Delta(r) =
	\begin{cases}
	\frac{1}{\delta_r}, &        d \le r \le d+\delta_r,\\
	0, &         r< d~ \text{or }~r>d+ \delta_r.
	\end{cases} 
	\label{Eq6}  
	\end{equation}
	In the Baxter sticky limit, $\delta_r \rightarrow 0$, $\epsilon \rightarrow \infty$, while the sticky length $l_g \equiv \delta_r(\mathrm{e}^{ \beta \epsilon} -1)$
	remains constant, the Boltzmann factor reduces to
	\begin{equation}  
	e^{-\beta u_{st}(r)} = 1+l_g \delta(r-d),
	\label{Eq5}
	\end{equation} 
	where $\delta(r)$ is the Dirac delta function. The sticky length accounts for the strength of covalent bonds between the atoms and will be directly related to the acid ionization constant.  
	
	The equation of state can be calculated using either the ``physical picture", which takes into account only ``atoms" $\ch{H}$ and $\ch{A}$, or an alternative ``chemical picture" in which besides the free unassociated particles $\ch{H}$ and $\ch{A}$ 
	there are also present molecules $\ch{HA}$.  Clearly both approaches must lead to the same equation of state~\cite{terrell}. Within the physical picture the osmotic
	pressure $P$ can be obtained using the virial expansion~\cite{mcquarrie}:
	\begin{equation}  
	\beta P = c_{\ch{H}} +c_{\ch{A}} +B_{\ch{HH}}c_{\ch{H}}^2 +B_{{\ch{AA}}}c_{\ch{A}}^2+2 B_{{\ch{HA}}} c_{\ch{H}} c_{\ch{A}}+\mathcal{O}(c_{\ch{A}}+c_{\ch{H}})^3,
	\label{Eq8}
	\end{equation}
	where  $B_{ij}$ are the second virial coefficients\cite{mcquarrie}:
	\begin{equation} 
	\begin{split} 
	B_{ij} = 2 \pi  \int_{0}^{\infty} (1-\mathrm{e}^{- \beta u_{ij}(r)}) r^2 dr .
	\label{Eq9}
	\end{split}
	\end{equation}
	For the case $i=j$ the interaction is just the hard sphere repulsion, so that $B_{AA} =B_{HH}= \frac{2 }{3}  \pi d^3$ and $B_{{\ch{HA}}} =  \frac{2}{3}\pi d^3 + \overline{B}_{{\ch{HA}}}$,
	where   $\overline{B}_{{\ch{HA}}}=2 \pi\int_{ d}^{\infty} (1-\mathrm{e}^{- \beta u_{ij}(r)})  r^2 dr$.
	On other hand, in the chemical picture there are three species: free $\ch{H}$ and $\ch{A}$, as well as molecules $\ch{HA}$. The equation of state can be written in terms of the respective concentrations designated by $c^*$, such that $c^*_{\ch{A}}=c_{\ch{A}}-c^*_{\ch{HA}}$ and
 $c^*_{\ch{H}}=c_{\ch{H}}-c^*_{\ch{HA}}$, and to second order in density is 
	\begin{equation}  
	\beta P =  c_{\ch{H}}^* + c_{\ch{A}}^* +c^*_{\ch{HA}} +\frac{2 \pi}{3}  d^3(c^*_{\ch{A}}+c^*_{\ch{H}})^2+
	\mathcal{O}(c^*_{\ch{A}}+c^*_{\ch{H}})^3
	\label{Eq11}
	\end{equation}
	where the fourth term is just the hardcore contribution to the osmotic pressure.  In equilibrium $c^*_{\ch{HA}}  = K_{eq}^0{c_{\ch{H}}^*} c_{\ch{A}}^*$, where $K_{eq}^0$ is the equilibrium constant. Comparing Eq.~\ref{Eq8} and Eq.~\ref{Eq11},  we obtain
	\begin{equation}  
	K_{eq}^0 = -2\overline{B}_{{\ch{HA}}}=4\pi \int_{d}^{\infty}(\mathrm{e}^{-\beta u_{st}(r)}-1)r^2 dr,
	\label{Eq12}
	\end{equation}
	In the Baxter sticky limit the equilibrium constant simplifies to
$K_{eq}^0 =4 \pi d^2 l_g$.
	In this simple calculation our particles H and A interacted only through a hard core repulsion and a short range attraction. If we are interested in modeling the acidic groups, both H and A must also carry charge, $\ch{H^+}$ and $\ch{A^-}$.  In this case the calculations become much more involved, since the usual virial expansion diverges and instead a certain class of diagrams must be summed together to obtain a convergent result~\cite{mcquarrie}.  This leads to a non-analytic term proportional to $c^{3/2}$ in the density expansion.  Falkenhagen and Ebeling ~\cite{Falkenhagen,LEVIN1996} studied this problem  in order to account for the formation of Bjerrum pairs in 1:1 electrolyte, and we can extend their results to the particles which in addition to the Coulomb force also interact through a short range sticky potential, for details of the derivation see Supplementary Information (SI).  In our case the equilibrium constant becomes
	\begin{equation}  
	K_{eq} = 4\pi d^2 l_g \mathrm{e}^{b} + K_{Eb},   
	\label{Eq14}
	\end{equation}
	where $b=\lambda_B/d$ and $\lambda_B$ is the Bjerrum length $q^ 2 /\epsilon_w k_B T$.   The Ebeling equilibrium constant is~\cite{eblingo68} 
	\begin{equation}  
	\begin{split}
	K_{Eb} = 8\pi d^3\big( \frac{1}{12}
	b^3\left(\mathrm{Ei}\left(b\right)-\mathrm{Ei}\left(-b\right)\right)-\frac{1}{3}\cosh{b}-\\ \frac{1}{6}b\sinh{b}-\frac{1}{6}b^2\cosh{b}+\frac{1}{3}+\frac{1}{2}b^2\big),
	\end{split}   
	\label{Eq15}
	\end{equation}
	where $\mathrm{Ei}$ is the exponential integral function.
	\begin{figure}
\vspace{-1cm}
		\begin{center}
			\includegraphics[width=6cm]{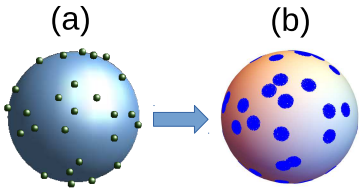}
		\end{center}
\vspace{-5cm}
		\caption{(a) Representation of a colloidal particle with spherical $\ch{A^-}$ sticky site on its surface.  (b) Mapping of spherical sticky sites onto disk-like surface patches used in the new theory of CR. \\
}  
		\label{fig2}
	\end{figure}

	We can now explore the validity of NP theory by constructing a simple model of a colloidal particle with $\ch{A^-}$ sticky surface groups, see Fig.\ref{fig2}a. This model can then be studied using Monte Carlo (MC) simulations~\cite{Frenkel,Allen}.  
	Knowledge of the equilibrium constant $K_{eq}$ will also allow us to directly compare the effective colloidal charge and the ionic density profiles obtained using NP theory with the results of MC simulations.
	
	The simulations are performed inside a spherical Wigner-Seitz (WS) cell of radius $R$, determined by the colloidal volume fraction in the suspension, $v=a^3/R^3$.
	A colloidal particle of radius $a=100$\AA$\,$ and $Z$ spherical adsorption sites of radius $2$\AA$\,$ and charge $-q$, 
	randomly distributed on its surface, is placed at the center of the simulation cell, see Fig. \ref{fig2}a. 
	The bare colloidal charge $-Zq$ is the same as the number of adsorption sites. The WS cell also contains  hydronium ions at bulk concentration $c_a=10^{-pH}$, derived from the dissociation of a strong acid, as well as 1:1 strong electrolyte at concentration $c_s$.  The hydronium ions interact with the adsorption sites through both Coulomb and short range Baxter potential with $l_g=109.97$\AA, while all other ions interact only through the Coulomb force. All ions and sticky sites are modeled as hard spheres of radius $r=2$\AA, with a point charge at the center. Note that the alkali metal cations derived from the dissociation of salt can come as close to the adsorption sites as the
hard core repulsion allows, however, since there is no chemical
		reaction between \ch{Na^+} and carboxyl in water  --- sodium acetate is a very strong
		electrolyte and is fully dissociated in water ---  we assume that there is only Coulomb interaction between alkali metal cations and the surface groups.
	\begin{figure}
		\begin{center}
			\includegraphics[width=7cm]{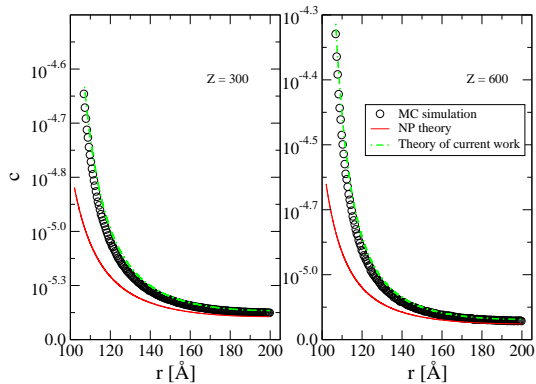}
		\end{center}
		\caption{Density profile of hydronium counterions measured in particles per \AA$^3$. Symbols are the simulation data and solids (red) and dashed (green) lines are the predictions of the NP theory and of the theory developed in the present Letter, respectively. The parameters are $a= 100$\AA, $R= 200$\AA, and $l_g=109.97$\AA. In a) and b) the colloidal particles have respectively $300$ and $600$ active surface sites. }  
		\label{fig1}
	\end{figure}

	  To perform the simulations we have used a progressively smaller values of $\delta_r$, and larger values of $\epsilon$ to 
	check the convergence to the Baxter sticky limit, see the discussion in SI. 
	The solvent is treated as a dielectric continuum 
	of permittivity $\epsilon_w = 80$, with Bjerrum length $\lambda_B = 7.2 $\AA.  If hydronium ion adsorbs to a sticky site, the site becomes inactive (stickiness is turned off) and no other hydroniums can be adsorbed.  This mimics the chemical reaction which takes place at the adsorption (sticky) site. Note that in the simulation the sites are active or inactive depending on whether there is a hydronium ion within the range of the potential $u(r)$, see SI for more details. 
	We have used $5 \times 10^6$ MC steps for equilibration and $10^5$ steps for production. 
	
In Fig.~\ref{fig1} we compare the density profiles obtained using the MC simulations with the predictions of NP theory using the bulk equilibrium constant derived in Eq. (\ref{Eq14}).   We see that there is a significant deviation between the theory and simulations, even when only neutralizing hydronium ions are present inside the simulation cell.     
	
	We can trace the breakdown of the NP theory to the use of the bulk equilibrium constant to account for the surface chemical reaction.  While the particles in the bulk are free to move, the adsorption sites are bound to the surface.  This affects the entropic contribution to the adsorption free energy.  Furthermore, as can be seen from Eq. (\ref{Eq14}),
	the bulk equilibrium constant also includes a contribution from the Coulomb interaction between the two oppositely charged ions that form a neutral molecules.  On the other hand the Coulomb interaction is also taken into account in the solution of the PB equation and is, therefore, counted twice.  Finally, the concentration of ions near a strongly charged surface can be so large that the use of concentration instead of activity, might not be justified. 
	In view of these observations, we now propose a different approach to CR.

	Let us first imagine that the whole colloidal surface is uniformly sticky. The concentration of ions around the colloidal particle will then satisfy a modified PB (mPB) equation
	\begin{equation}  
	\nabla^2 \phi(r) =  -\frac{4 \pi q}{\epsilon_w} \left[-\sigma_0 \delta(r-a) + c_{\ch{H^+}}(r) +c_+(r)- c_-(r) \right] ,
	\label{Eq17}
	\end{equation}   
	where $\sigma_0=Z/4 \pi (a+r_{ion})^2$.  The ionic concentrations are:
	\begin{eqnarray}
	&&c_{\ch{H^+}}(r) = c_a~\mathrm{e}^{-\beta \left( u(r) +q \phi(r)\right)}\\ 
	&& c_+(r)= c_s~\mathrm{e}^{-\beta q \phi(r)}\\
	&&c_-(r) = (c_a+c_s)~\mathrm{e}^{ \beta q \phi(r)},	
	\label{Eq18}
	\end{eqnarray}  
	where the bulk concentration of hydronium is $c_a=10^{-pH}$, and $c_s$ is the bulk concentration of salt.

	Using Eq.~\ref{Eq5} we obtain $\mathrm{e}^{-\beta \left( u(r) +q \phi(r)\right)}=[1-l_g \delta(r-a)]\mathrm{e}^{-\beta q \phi(r)}$.
	The surface concentration of adsorbed ions is then
	\begin{equation}
	\begin{split}  
	\sigma_a =c_a l_g \mathrm{e}^{-\beta q \phi_0},
	\end{split}
	\label{Eq18a}
	\end{equation}  
	where $\phi_0=\phi(a+r_{ion})$ is the contact electrostatic potential. The net surface charge density is then 
	$q \sigma_{net}=-q \sigma_0+q \sigma_a$.
	
\begin{figure}
		\begin{center}
			\includegraphics[width=8cm]{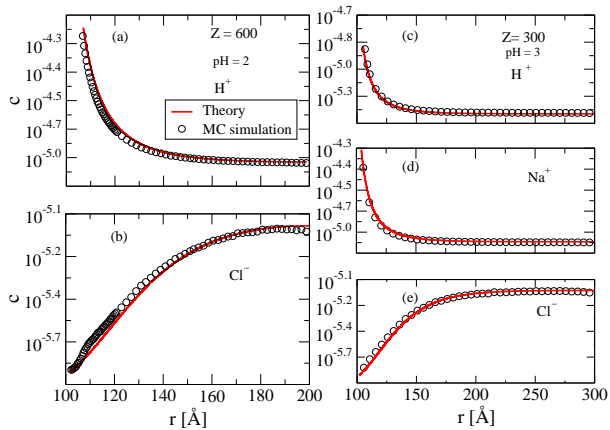}
		\end{center}
		\caption{Comparison between theory (symbols) and simulation (solid lines).   Panel (a) is the density profile of hydronium and (b) of $\ch{Cl^-}$  for colloidal particles with  $Z=600$ functional groups and volume fraction of $12.5$\% in hydrochloric acid at pH$=2$.  Panels (c), (d) and (e) are the hydronium, $\ch{Na^+}$, and $\ch{Cl^-}$, density profiles respectively, for colloidal particles with $Z=300$ functional groups and volume fraction of  $3.7$\%, in a solution of hydrochloric acid of pH$=3$ and 10mM of \ch{NaCl}. The density $C$ is in units of particles per \AA$^3$. 
The sticky length is $l_g=109.97$\AA.\\
			\vspace{0.5cm}  }
		\label{fig3}
	\end{figure}

	The real colloidal surface, however,
	is not uniformly sticky, instead hydronium ions can adsorb only at the  specific functional groups, 
	see, Fig.~\ref{fig2}a. To understand the role of discreteness of the surface charge~\cite{jho2012}, let us first consider a much simpler problem.  Suppose we have an isolated colloidal particle (no other ions), with $Z$ fully ionized surface groups.   How much work must be done to bring a counterion from infinity to the contact with one of the surface groups? To answer this question, let us first consider a spherical 2d one component plasma (OCP) of $Z$ charged point particles on a sphere of radius $a$ with a {\it neuralizing} uniform background. In a crystal or amorphous state the electostatic energy of this OCP is  $F^{OCP} \approx -Mq^2 Z^{3/2}/2 \epsilon_w a$, where $M=1.106$ is the Madelung constant~\cite{levin}.  On the other hand the OCP energy can also be split into distinct contributions: $F^{OCP}=Z^2 q^2/2 \epsilon_w a -Z^2 q^2/\epsilon_w a +F^{qq}$, where the first term is the self energy of the neutralizing background, the second term is the interaction of the discrete charges with the background, and the last term is the interaction energy between the discrete charges. This last term is of particular interest to us since it is precisely the electrostatic energy of an isolated colloidal particle with {\it discrete} surface groups. Using the expressions above, it can be written as:
	\begin{equation}
	F^{qq}(Z)=\frac{Z^2 q^2}{2 \epsilon_w a}-\frac{M q^2 Z^{3/2}}{2\epsilon_w a} .
	\end{equation}
	The work required to bringing a counterion to the colloidal surface equals to the change in the electrostatic energy 
$\mu=F^{qq}(Z-1)-F^{qq}(Z)$ or, 
	\begin{equation}
	\mu \approx -\frac{\partial F^{qq}}{\partial Z}=-\frac{Zq^2}{\epsilon_w a}+\frac{3 Mq^2 Z^{1/2}}{4\epsilon_w a} .
\label{mu}
	\end{equation}
	Note that the first term on the right of Eq. (\ref{mu}) is just the usual mean-field interaction energy between a counterion and a uniformly charged sphere $q\phi_0$, where  $\phi_0$ is the ``mean-field" surface potential, while the second term is the correction due to discrete nature of the surface charge groups.  If the counterion is brought into contact with one of the surface groups, the total work $\varphi_0$ is
	\begin{equation}
	\beta \varphi_0=\beta q\phi_0+\frac{3 M \lambda_B Z^{1/2}}{4a}-\frac{\lambda_B}{d} ,
	\label{var}
	\end{equation}
	where the last term is the direct energy of interaction between the site and the adsorbed counterion. With these insights, we now return to the problem of a colloidal particle inside an electrolyte solution.
	To simplify the geometry we will map the sticky spherical sites onto sticky circular disk 
	patches of the same effective contact area.  Due to hardcore repulsion only half of the area of a spherical sticky site is available for adsorption, the patches must then have radius $r_{patch} = \sqrt{2}~d$, see Fig~\ref{fig2}b. Once adsorption takes place, the site becomes inactive, but continues to interact with the other ions through the Coulomb potential. The fact that only part of the colloidal surface is sticky can be taken into account by the renormalization of the sticky length  $l_g\longrightarrow l_g^{eff}=l_g \alpha_{eff}$, where $\alpha_{eff}$ is the fraction of colloidal surface area occupied by the {\it active} sticky patches,
	\begin{equation}
	\alpha_{eff} = \frac{N_{site}^{act}r_{patch}^2}{ 4 a^2}.  
	\label{Eq19}
	\end{equation}  
	The number of {\it active} sites, $N_{site}^{act}$, is determined from Eq.(\ref{Eq18a}) with $l_g\longrightarrow l_g^{eff}$ and $\phi_0 \rightarrow \varphi_0$, so that $N_{site}^{act} = Z -4 \pi a^2 c_a l_g^{eff}~\mathrm{e}^{-\beta \varphi_0}$.
	The surface electrostatic potential $\varphi_0$ is given by Eq. (\ref{var}), with $\phi_0$ now being the {\it mean-field} surface electrostatic potential, which must be calculated self consistently from the solution of PB equation.  Eq. $l_g^{eff}=l_g \alpha_{eff}$ and Eq. (\ref{Eq19}) can now be solved to obtain the effective sticky length
	\begin{equation}  
	l_{g}^{eff} =\frac{l_g Z r_{patch}^2}{4  a^2(1+ l_g c_a \mathrm{e}^{-\beta \varphi_0}\pi r_{patch}^2)}.
	\label{Eq21}
	\end{equation}
	The effective surface charge density to be used as the boundary condition for PB equation is then $q \sigma_{eff} = -q\sigma_0 +q l_g^{eff} c_a \mathrm{e}^{-\beta \varphi_0}$.
	We now solve the PB equation with the boundary conditions  $\phi'(a+r_{ion})=4\pi q \sigma_{eff}/\epsilon_w$ and $\phi'(R)=0$, due to the overall charge neutrality.
The calculation is performed numerically using the 4th order Runge-Kutta,
in which the value of the surface potential $\phi(a+r_{ion})=\phi_0$
is adjusted based on the Newton-Raphson algorithm to obtain zero electric field at the cell boundary.
In Fig.~\ref{fig3} we compare the ionic density profile obtained using MC simulations and the new theory. The agreement is excellent, without any adjustable parameters.

\begin{figure}
		\begin{center}
			\includegraphics[width=5cm]{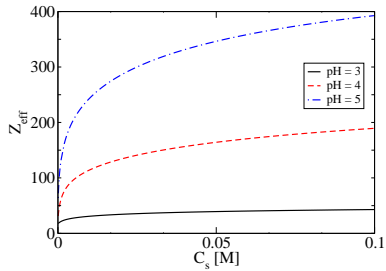}
		\end{center}
		\caption{Effective charge of a colloidal particle of radius $a=100$\AA$\,$ and $Z=600$ carboxyl groups, as a function of salt concentration, in a solution of a given pH obtained using the new theory of CR. Colloidal volume fraction is $12.5$\%.}  
		\label{fig4}
	\end{figure}

Having established the accuracy of the theoretical approach, we now use it to 
	calculate the effective charge of colloidal particles stabilized by surface carboxyl groups with acid 
	ionization constant $K_a=1.8\times 10^{-5}$M.
	Note that the equilibrium constant $K_{eq}$ defined in the present work is the inverse of $K_a$. Using $K_{eq}=1/K_a$ in Eq. (\ref{Eq14}) allows us to obtain the sticky length $l_g$. In Fig. \ref{fig4} we show the dependence of the effective colloidal charge  on the pH and salt concentration for particles with $Z=600$ functional groups on the surface.  In SI we also plot the behavior of the modulus of the contact electrostatic potential.  While $Z_{eff}$ increases with salt, the modulus of the contact potential and, therefore, the zeta potential decrease with the salt concentration~\cite{Borkovec_ex}. 
	
	In this Letter we have presented a model of colloidal particle with sticky adsorption sites. Analyzing the thermodynamics
	of ionic association, we were able to relate the interaction potential between the adsorption sites and hydronium ions with the bulk equilibrium constant.
With the help of this model we discovered that existing approaches were not able to quantitatively account for CR, predicting incorrect value of colloidal charge and ionic density profiles which deviated significantly from simulations.  With the insights gained from the simulations, we were able to introduce a new theory of CR. 
The next step is to explore the role of CR on the interaction between colloidal particles and the role that it will play in stability of colloidal suspensions. This should be possible to do by implementing the CR boundary
condition analogously to the approach recently employed for metal particles~\cite{santos2019}. 
	
	This work was partially supported by CNPq, CAPES, and  the US-AFOSR under the grant FA9550-12-1-0438.

\bibliography{ref}
\end{document}


\author{Amin Bakhshandeh}
	\email{amin.bakhshandeh@ufrgs.br}
	\affiliation{Programa de Pos-Gradua\c c\~  ao em F\'\i sica, Instituto de F\'\i sica e Matem\'{a}tica, Universidade Federal de Pelotas, Caixa Postal 354, CEP 96010-900 Pelotas, RS, Brazil}
	\author{Derek Frydel}
	\email{derek.frydel@usm.cl}
	\affiliation{Department of Chemistry, Federico Santa Maria Technical University, Campus San Joaquin, Santiago, Chile}
	\author{Alexandre Diehl} 
	\email{diehl@ufpel.edu.br}
	\affiliation{Departamento de F\'\i sica, Instituto de F\'\i sica e Matem\'{a}tica, Universidade Federal de Pelotas, Caixa Postal 354, CEP 96010-900 Pelotas, RS, Brazil}
	\author{Yan Levin}
	\email{levin@if.ufrgs.br }
	\affiliation{Instituto de F\'isica, Universidade Federal do Rio Grande do Sul, Caixa Postal 15051, CEP 91501-970, Porto Alegre, RS, Brazil}

	{\Large{\bf Supplementary Material} } \\[0.5cm]
	\section{derivation of Eq.(9)}
	In the Letter we have derived the equilibrium constant for uncharged sticky particles.  The equilibrium constant in our model is exact and does not depend on the density of the reactants.  The higher terms of the virial expansion will modify the activity coefficients, so that in the law of mass-action the concentration will have to be replaced by activity. However since the Poisson-Boltzmann equation does not take into account ionic
	correlations, to remain consistent with PB, the activity coefficients must be set to unity. In other words, activity in our theory is equal to the concentration and the higher order terms within the virial expansion must be neglected.
	
	Here we present a physically intuitive derivation of the equilibrium constant for sticky ions, Eq.(9). In the case of purely electrostatic interactions, the equilibrium constant was first 
	derived by  
	Ebeling~\cite{Ebeling} considering the exact density expansion of the equation of state up to $\rho^{5/2}$ order.  The original treatment of Ebeling and Falkenhagen~\cite{Falkenhagen} is quite involved, however, the final expression for the equilibrium constant has a simple physical interpretation. The Ebeling equilibrium constant is   
	\begin{equation}  
	K_{Eb} = 8\pi d^3\left\{ \frac{1}{12}
	b^3\left[\mathrm{Ei}\left(b\right)-\mathrm{Ei}\left(-b\right)\right]-\frac{1}{3}\cosh{b}-\\ \frac{1}{6}b\sinh{b}-\frac{1}{6}b^2\cosh{b}+\frac{1}{3}+\frac{1}{2}b^2\right\}
	\label{Eq13}
	\end{equation}
	In the strong coupling limit, large $b$ values, the equilibrium constant has an asymptotic series expansion  
	\begin{equation}  
	K_{Eb} = 4 \pi d^3 \frac{\mathrm{e}^b}{b} \left(1+\frac{4}{b}+\frac{4\times 5}{b^2}+\frac{4\times 5 \times 6}{b^3}+...\right) \,.
	\label{Eq14}
	\end{equation}
	This is exactly the same as the asymptotic expansion of the Bjerrum equilibrium constant defined as an internal partition function of a ``molecule" formed by a cation-anion pair: 
	\begin{equation}  
	K_{Bj} = 4\pi \int_{d}^{R_{Bj}}\mathrm{e}^{ \frac{\lambda_B}{r}}r^2 dr,
	\label{Eq15}
	\end{equation}
	The Bjerrum cutoff is defined as $R_{Bj}=\lambda_B/2$.  However, in the strong coupling limit, the value of the equilibrium constant is insensitive to the value of the cutoff~\cite{levin}. 
	On the other hand, in the weak coupling limit, Bjerrum constant will have a strong cutoff dependence and is not well defined.  The Ebeling equilibrium constant can be considered as an analytical continuation of the Bjerrum equilibrium constant between weak and strong coupling limits.   
	
	With these insights, we now turn to the sticky electrolytes, which besides the Coulomb interaction have a sticky short range potential between cations and anions.  This sticky potential models the chemical
	(quantum) bond that can form between carboxyl and hydronium.    
	The Bjerrum equilibrium constant for sticky cation-anion cluster can be written as:
	\begin{equation}  
	K_{eq} = 4\pi \int_{d}^{R_{Bj}}\mathrm{e}^{-\beta u_{st}(r)+ \frac{\lambda_B}{r}} r^2 dr.
	\label{Eq16}
	\end{equation}
	Using Eq. 4 of the paper we can write this as
	\begin{equation}  
	K_{eq} = 4\pi \int_{d}^{R_{Bj}}\left[(1+l_g \delta(r-d)\right]\mathrm{e}^{ \frac{\lambda_B}{r}}r^2 dr,
	\label{Eq17}
	\end{equation}
	After integrating the delta function we obtain
	\begin{equation}  
	K_{eq} = 4\pi d^2 l_g \mathrm{e}^{b}+\int_{d}^{R_{Bj}} \mathrm{e}^{ \frac{\lambda_B}{r}}r^2 dr,
	\label{Eq18}
	\end{equation}
	where the second term is just the Bjerrum equilibrium constant for a non-sticky electrolyte.  To extend the validity of this expression beyond the strong coupling limit, we replace 
	$K_{Bj}$ by $K_{Eb}$, resulting in Eq. 9 of the paper. 

For weak acids the first term on the right hand side of Eq. \ref{Eq18} will dominate the second term.  In this case, the theory presented in the Letter will reduce to the NP approach, Eq. (1) of the Letter, but with the bulk equilibrium constant replaced by the
effective surface equilibrium constant, $K_{eq} \rightarrow K_s$
\begin{equation}  
	K_{s} = \frac{K_{eq}}{2}\mathrm{e}^{-\frac{3 M \lambda_B Z^{1/2}}{4a}} \,.
	\label{Eq19}
\end{equation}

	
	\section{Convergence to the Baxter limit}
	
	Fig.1 shows a schematic representation of the colloidal surface with sticky sites that can interact with hydronium ions through a square well potential of depth $-\epsilon$ and width $\delta_r$.  The simulations are conducted for progressively decreasing values of $\delta_r$, and increasing $\epsilon$, while the sticky length $l_g=109.97$~\AA$\,$ is held fixed. 
	Only one hydronium ion can adsorb to a sticky site. After one hydronium is adsorbed, the $u_{st}$ interaction of the site with the other hydroniums is switched off, while the Coulomb force remains active.  This is due to the fact that we are modeling a monovalent acidic group such as carboxyl. For such functional groups only one hydronium ion per site can chemically react. We should also mention that we do not put any restrictions on the metal cations --- they can come as close to the  adsorption sites as the hard core constraint allows. However, since there is no chemical reaction between metal cations and carboxyl --- for example sodium acetate is a very strong electrolyte --- these ions do not have $u_{st}$ interaction with the adsorption sites, only the Coulomb interaction acts between the two. 
	Fig. 2 shows a quick convergence to the Baxter sticky limit.
	
	\begin{figure}
		\includegraphics[width=6cm]{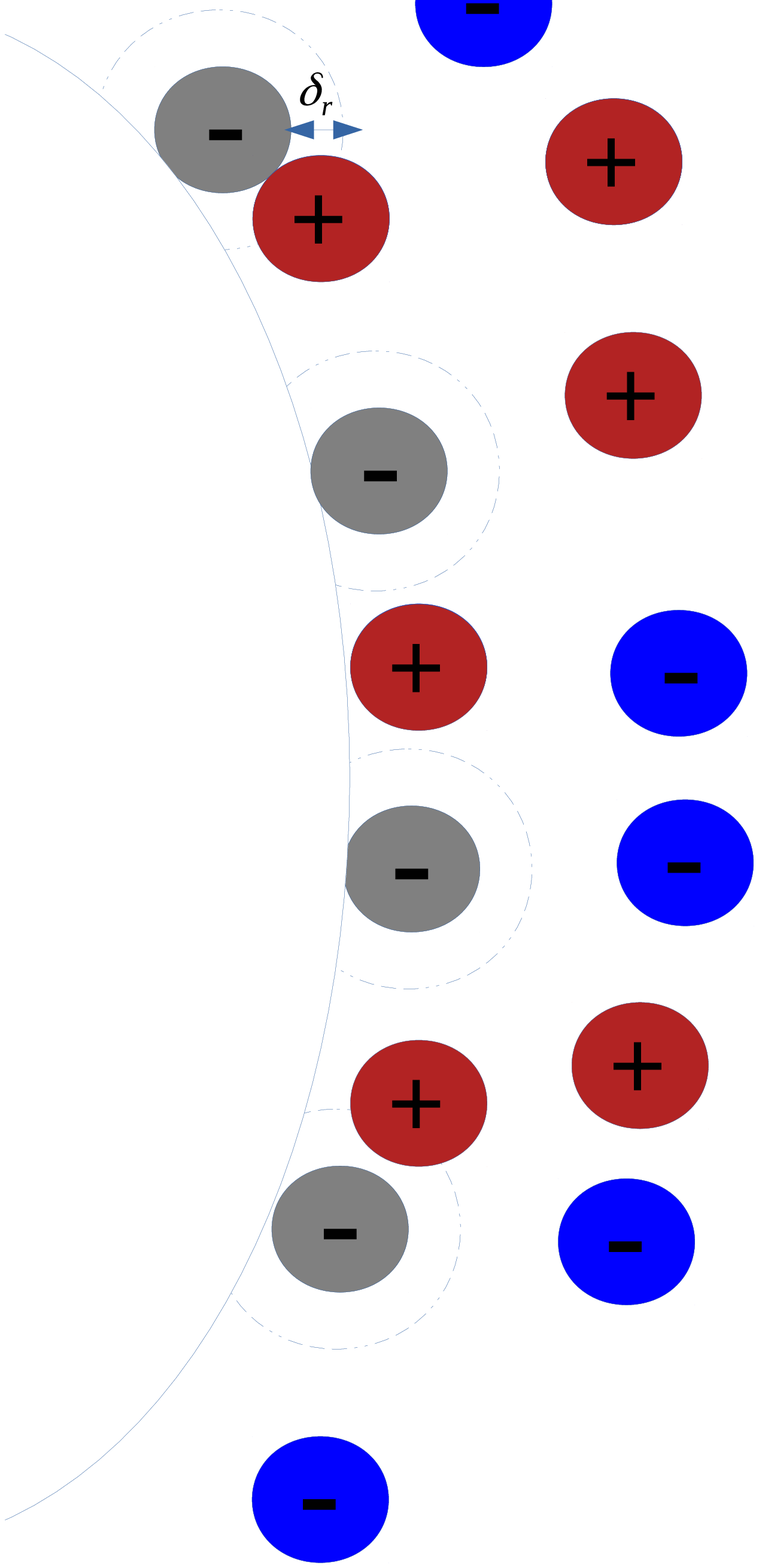}
		\caption{Schematic representation of Baxter sticky sites interacting with the hydronium ions through a square well potential of width $\delta_r$. \\
		}  
		\label{fig1}
	\end{figure}
	
	\begin{figure}
		\includegraphics[width=12cm]{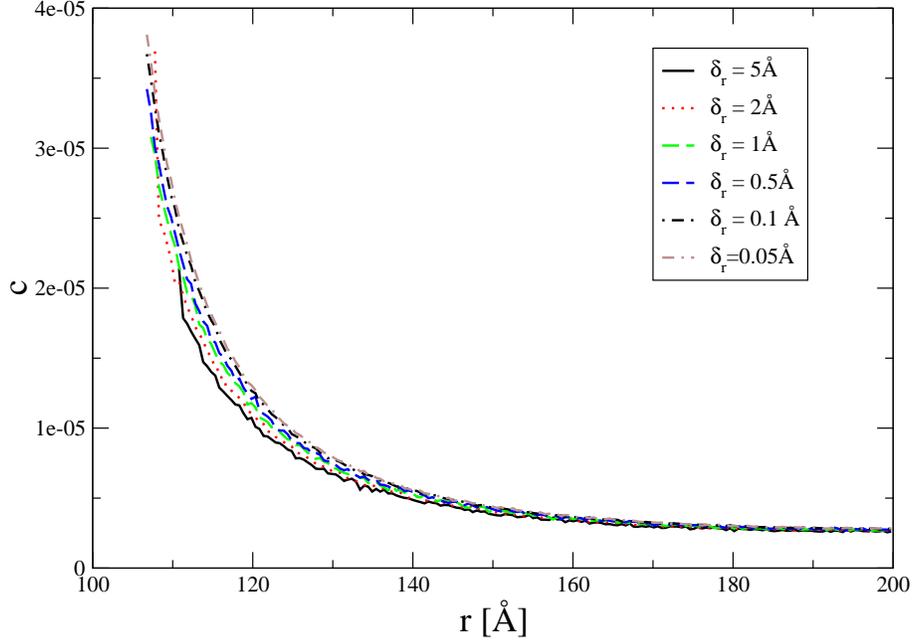}
		\caption{Density profile of ions for different size of $\delta_r$.  The density is in terms of the number of particles per \AA$^3$ \\
		}  
		\label{fig2}
	\end{figure}
	
	\section{Comparison of Ninham-Parsegian (NP) theory and the present approach}
	
	In Fig. 3 we show the comparison of NP theory and of the approach developed in the Letter with the MC simulations. We see that the deviations become stronger for larger surface charge.  
	\begin{figure}
		\includegraphics[width=12cm]{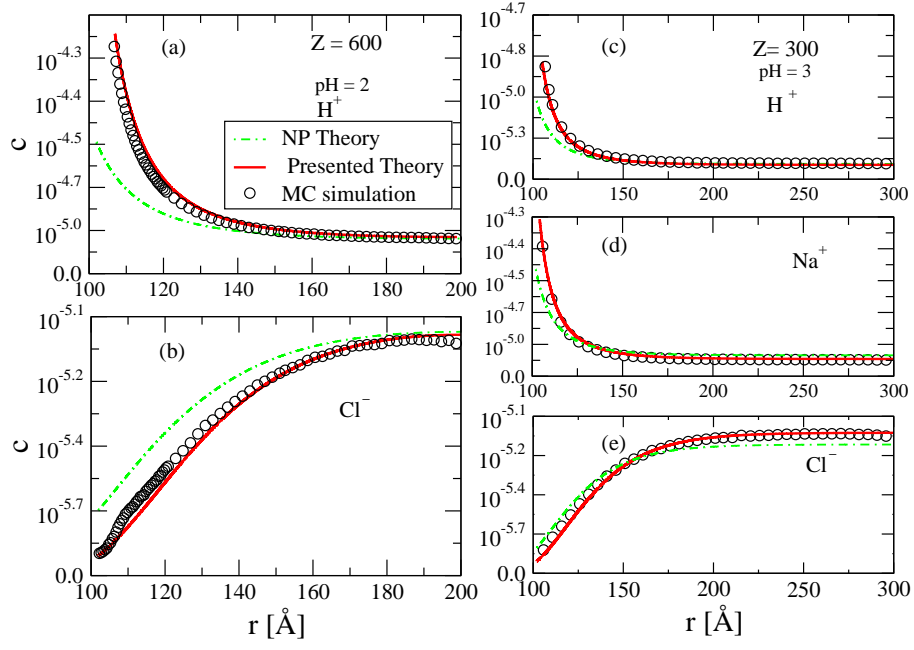}
		\caption{Density profile obtained using MC simulation compared with the two theories. Same parameters as in Fig. 3 of the Letter.\\
		}  
		\label{fig3}
	\end{figure}
In Fig. 4 we explore the deviation between the two theories for larger values of $Z$ and $l_g$, for which the simulations take impractically long time to equilibrate.  As expected, the deviation between the two approaches increase dramatically with $Z$ and $l_g$.
\begin{figure}
	\includegraphics[width=12cm]{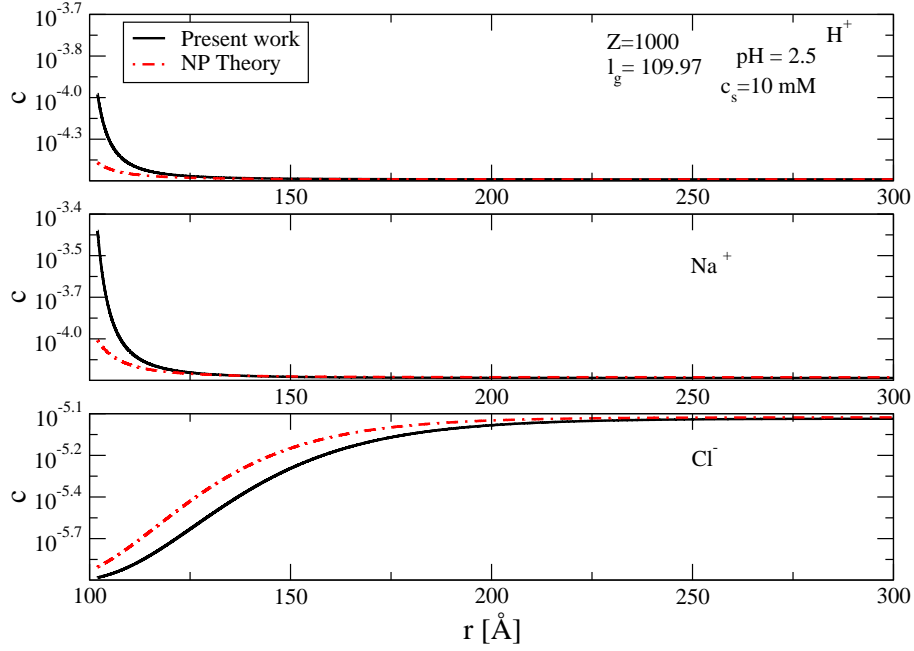}	\\
\vspace{1.5cm}
		\includegraphics[width=12cm]{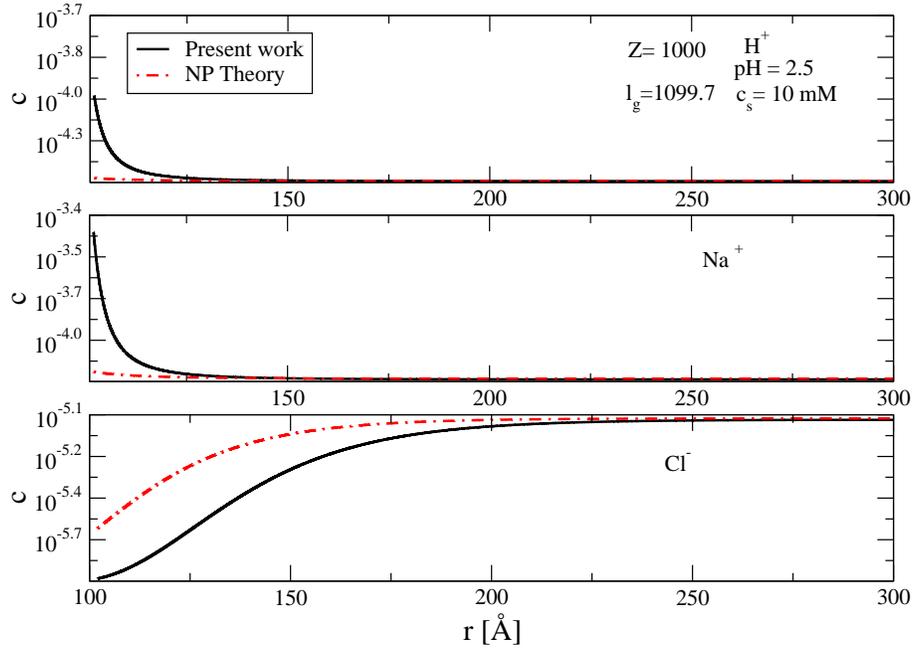}
		\caption{Comparison between the density profiles predicted by the theory presented in the Letter and the NP approach for two different values of $l_g$.\\
		}   
		\label{fig3a}
	\end{figure}

	\section{Contact Potential}
	In the Letter we showed that the effective colloidal charge increases as a function of salt concentration and pH.  In the experiments, however, zeta potential is more easily available than the effective charge.  Zeta potential, however, requires knowledge of the position of the slip plain, which depends on the surface chemistry.  Qualitatively, however we expect that the zeta potential will behave similarly to the modulus of the electrostatic contact surface potential.  Below we show the behavior of the modulus of the surface potential as a function of pH and salt concentration.
\begin{figure}
\vspace{1cm}
		\includegraphics[width=12cm]{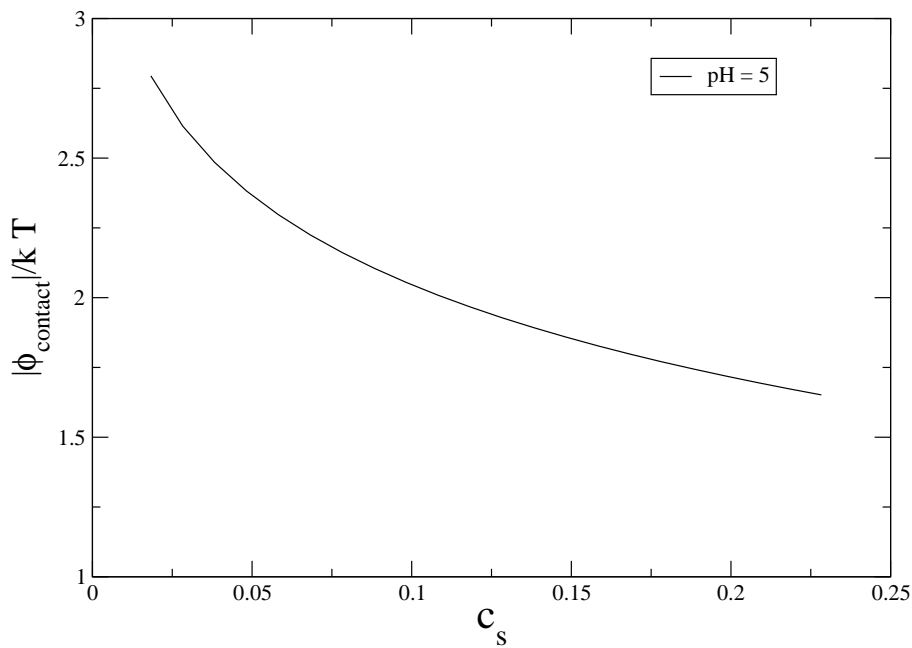}
		\caption{Modulus of the contact potential as a function of salt concentration. The colloidal particle has $600$  acetic acid functional groups and the pH is fixed at 5.  \\
		}  
		\label{fig4}
	\end{figure}
	\begin{figure}
		\includegraphics[width=12cm]{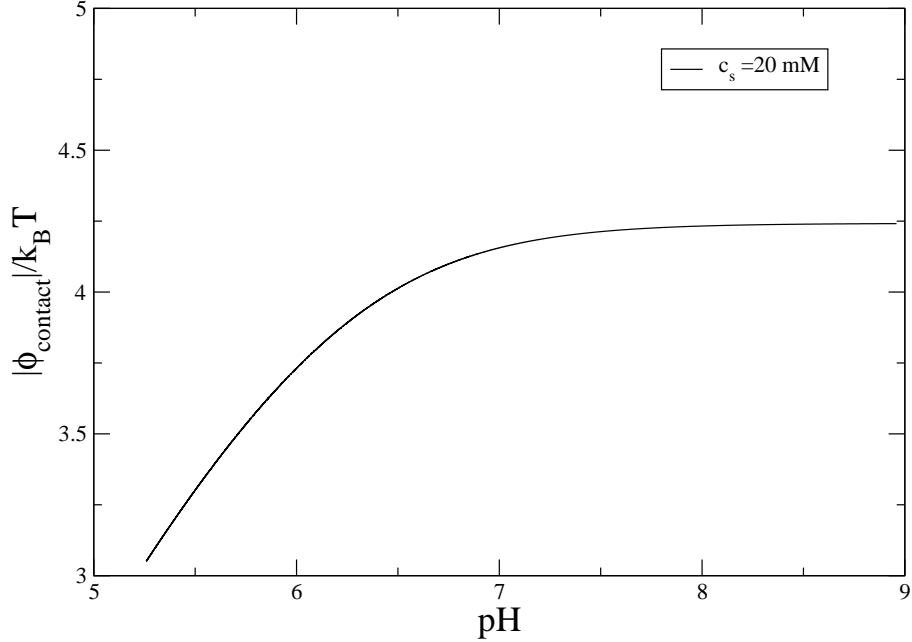}
		\caption{Modulus of the contact potential as a function of pH.   The colloidal particle has $600$ acetic acid functional group on it surface.  Salt concentration is fixed at 20 mM. \\
		}  
		\label{fig5}
	\end{figure}

Fig.~\ref{fig4} shows that the modulus of the contact potential decreases with increasing salt concentration, while Fig.~\ref{fig5} demonstrates that the modulus of the potential increases with the pH, similar to the behavior observed for zeta potential in experiments~\cite{Borkovec_ex}.
There is an opposite behavior of the zeta potential and the effective surface charge as a function of salt concentration. While the effective surface charge increases, zeta potential decreases. The reason for this is that the contact potential is negative, therefore, addition of salt screens the electrostatic interactions thus  {\it raising} the contact potential, i.e. it becomes less negative. This, in turn, decreases the concentration of hydronium ions near the surface, diminishing  their association with the  carboxyl groups, resulting in larger $Z_{eff}$ with increasing salt concentration.